\documentclass{PoS}

\usepackage{axodraw}

\usepackage{epsfig}

\def\beq{\begin{equation}}
\def\eeq{\end{equation}}
\def\beqa{\begin{eqnarray}}
\def\eeqa{\end{eqnarray}}

\title{Higher order corrections to $H^{\pm}$ production}

\ShortTitle{Higher order corrections to $H^{\pm}$ production}

\author{\speaker{Nikolaos Kidonakis}%
        \thanks{This work was supported by the National Science 
Foundation under Grant No. PHY 0555372.}\\
       Kennesaw State University, USA\\
       E-mail: \email{nkidonak@kennesaw.edu}}

\abstract{I discuss recent calculations of higher-order corrections 
to charged Higgs production through various partonic subprocesses. 
Particular attention is paid to $H^-$ production in association with a top
quark at the LHC.}

\FullConference{Prospects for Charged Higgs Discovery at Colliders\\
         16-19 September 2008\\
         Uppsala, Sweden}

\begin{document}

\section{Introduction}

Many models for new physics involve a more complicated Higgs sector than 
in the Standard Model. 
In the Minimal Supersymmetric Standard Model (MSSM) and other  
two-Higgs-doublet models (2HDM), one Higgs doublet gives mass to the 
up-type fermions and the other to the down-type fermions, with  
$\tan \beta=v_2/v_1$ the ratio of the vacuum expectation values 
for the two doublets.
The five physical Higgs particles in the MSSM include a light scalar,   
$h^0$, a heavy scalar, $H^0$, a pseudoscalar, $A^0$, and two charged Higgs 
bosons, $H^+$ and $H^-$.
A future discovery of a charged Higgs boson would constitute a definite sign 
of new physics. The Large Hadron Collider (LHC) is well positioned 
for a discovery of a charged Higgs \cite{Higgs}.

A lot of work on higher-order QCD and SUSY corrections to 
charged Higgs production 
has been performed over the last several years, 
including next-to-leading order (NLO) 
calculations for $bg \rightarrow tH^-$ 
\cite{Zhu01,Plehn02,BHJP03}, 
$b{\bar b} \rightarrow H^+ W^-$ \cite{HolZhu01,GLL07},
$b{\bar b} \rightarrow H^+ H^-$ \cite{Hou05,AlPl05}, 
$q{\bar q} \rightarrow H^+ H^-$ \cite{AlPl05}, 
as well as results for higher-order soft-gluon corrections 
for $bg \rightarrow tH^-$ \cite{NKcH,NKNNNLO}.

\section{Associated $H^-$ and top quark production}

We begin with the dominant process at the LHC, which is associated 
charged Higgs and top quark production. 
The leading order (LO) process is 
$bg \rightarrow t H^-$ and the corresponding 
Feynman diagrams are shown in Fig. {\ref{bgtH}.
The LO cross section is proportional to $\alpha \alpha_s
(m_b^2\tan^2 \beta+m_t^2 \cot^2 \beta)$
where $m_b$ is the bottom quark mass and $m_t$ is the top quark mass.

Yukawa and SUSY electroweak corrections for this process were calculated 
in \cite{JLOZ} and 1-loop SUSY corrections in \cite{BGGS,GLXY}.
The complete NLO QCD corrections were calculated 
in Ref. \cite{Zhu01,Plehn02,BHJP03}. 
The QCD corrections were shown to be substantial, contributing up to
85$\%$ enhancement of the lowest order cross section \cite{Zhu01}, 
and to  
reduce the scale dependence of the cross section.
The NLO SUSY-QCD corrections are smaller in comparison, 
with their precise value depending on MSSM parameters \cite{Plehn02,BHJP03}.

\begin{figure}
\begin{center}
\begin{picture}(80,120)(0,0)
\ArrowLine(0,75)(20,50)
\Gluon(0,25)(20,50){3}{5}
\ArrowLine(20,50)(60,50)
\DashLine(60,50)(80,25){3}
\ArrowLine(60,50)(80,75)
\Text(0,15)[c]{$g$}\Text(0,85)[c]{$b$}
\Text(40,65)[c]{$b$}
\Text(80,15)[c]{$H^-$}\Text(80,85)[c]{$t$}
\end{picture}
\hspace{30mm}
\begin{picture}(80,120)(0,0)
\ArrowLine(0,75)(40,75)
\Gluon(0,25)(40,25){3}{5}
\ArrowLine(40,75)(40,25)
\DashLine(40,75)(80,75){3}
\ArrowLine(40,25)(80,25)
\Text(0,15)[c]{$g$}\Text(0,85)[c]{$b$}
\Text(33,50)[c]{$t$}
\Text(80,15)[c]{$t$}\Text(80,85)[c]{$H^-$}
\end{picture}
\caption{LO diagrams for $bg \rightarrow t H^-$.}
\label{bgtH}
\end{center}
\end{figure}

To calculate the NLO QCD corrections, we have to include the one-loop 
virtual corrections to $bg \rightarrow t H^-$ and also the 
processes with one additional parton:

$bg \rightarrow t H^- g \quad  \quad
gg \rightarrow t H^- {\bar b} \quad \quad 
q {\bar q} \rightarrow t H^- {\bar b} \quad \quad
\quad b q \rightarrow t H^- q \quad \quad  
b {\bar q} \rightarrow t H^- {\bar q}$

$b b \rightarrow t H^- b \quad \quad 
b {\bar b} \rightarrow t H^- {\bar b}$

Issues with the calculation include the treatment of the 
bottom parton distribution, 
with a gluon splitting to $b{\bar b}$ in the collinear approximation,
valid for small $b$-quark $p_T$.
The diagrams for the process $gg \rightarrow {\bar b} t H^-$
with a gluon splitting to $b{\bar b}$ 
are shown in Fig. \ref{ggbtH}.

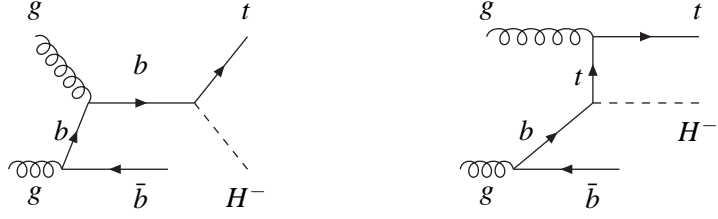
\begin{figure}
\begin{center}
\begin{picture}(80,120)(0,0)
\Gluon(0,75)(20,50){3}{5}
\ArrowLine(10,25)(20,50)
\Gluon(-10,25)(10,25){3}{3}
\ArrowLine(50,25)(10,25)
\ArrowLine(20,50)(60,50)
\DashLine(60,50)(80,25){3}
\ArrowLine(60,50)(80,75)
\Text(0,15)[c]{$g$}\Text(0,85)[c]{$g$}
\Text(10,40)[c]{$b$}
\Text(40,65)[c]{$b$}\Text(40,15)[c]{${\bar b}$}
\Text(80,15)[c]{$H^-$}\Text(80,85)[c]{$t$}
\end{picture}
\hspace{30mm}
\begin{picture}(80,120)(0,0)
\Gluon(0,75)(40,75){3}{5}
\ArrowLine(40,75)(80,75)
\ArrowLine(10,25)(40,50)
\Gluon(-10,25)(10,25){3}{3}
\ArrowLine(40,50)(40,75)
\ArrowLine(50,25)(10,25)
\DashLine(40,50)(80,50){3}
\Text(0,15)[c]{$g$}\Text(0,85)[c]{$g$}
\Text(40,15)[c]{${\bar b}$}
\Text(15,40)[c]{$b$}\Text(35,60)[c]{$t$}
\Text(80,40)[c]{$H^-$}\Text(80,85)[c]{$t$}
\end{picture}
\caption{LO diagrams for $gg \rightarrow {\bar b} t H^-$ with 
a gluon splitting to $b{\bar b}$.}
\label{ggbtH}
\end{center}
\end{figure}

Work on matching the processes $b g \rightarrow t H^-$ and 
$gg \rightarrow {\bar b} t H^-$  \cite{JFG} 
has been performed in \cite{BKP,MDPR,AlRa04}.
The use of matrix elements at large $p_T$ and parton showers 
at small $p_T$, results in double counting for small $p_T$ when 
doing a simple addition.
The matching performed in \cite{AlRa04} involves an analytic 
double-counting subtraction term, 
and it can be implemented in event generators to give 
smooth differential distributions for all $b$-quark $p_T$.

\subsection{$bg \rightarrow tH^-$ near threshold}

Higher-order corrections to charged Higgs production near threshold 
have been calculated at next-to-next-to-leading order (NNLO) in 
Ref. \cite{NKcH} and at next-to-next-to-next-to-leading order 
(NNNLO) in Ref. \cite{NKNNNLO}.

For the process 
$b(p_b) + g(p_g) \longrightarrow t(p_t)+H^-(p_{H})$ 
we define the kinematical invariants 
$s=(p_b+p_g)^2$, $t=(p_b-p_t)^2$, $u=(p_g-p_t)^2$,
and $s_4=s+t+u-m_t^2-{m_{H}}^2$, where $m_t$ is the top quark mass 
and $m_H$ is the charged Higgs mass.
Note that, while we use the $\overline{\rm MS}$ bottom quark mass  
in the coupling, we set $m_b=0$ in the kinematics.
At threshold $s_4 \rightarrow 0$, and the  
soft-gluon corrections take the form $[\ln^l(s_4/m_{H}^2)/s_4]_+$} 
and can be resummed.
For the order $\alpha_s^n$ corrections $l \le 2n-1$.
The leading logarithms (LL) are with  $l=2n-1$ and the 
next-to-leading logarithms (NLL) are with $l=2n-2$.
 
Near threshold these soft-gluon  corrections are dominant and 
provide good approximations to the complete QCD corrections. 
The NLO and NNLO soft-gluon corrections were calculated at NLL accuracy 
in \cite{NKcH}. Furthermore, the NNNLO soft NLL corrections were 
presented in \cite{NKNNNLO}.
 
The calculation of these corrections is derived from the fixed-order 
expansion of the resummed cross section.
Resummation follows from factorization properties of the
cross section and is performed in moment space. 
We can write the resummed cross section as \cite{NKcH,NKNNNLO,NNLOuni}
\beqa
{\hat{\sigma}}^{res}_{bg \rightarrow t H^-}(N) &=&
\exp\left[ \sum_i E_i(N_i)\right] \;
\exp \left[\sum_i 2\int_{\mu_F}^{\sqrt{s}} \frac{d\mu}{\mu}\;
\gamma_{i/i}\left(N_i,\alpha_s(\mu)\right)\right] \;
\exp \left[\sum_i 2 \int_{\mu_R}^{\sqrt{s}} \frac{d\mu}{\mu}\;
\beta\left(\alpha_s(\mu)\right)\right] \;
\nonumber\\ &&  \hspace{-10mm} \times \,
H^{bg \rightarrow t H^-}\left(\alpha_s(\mu_R)\right) \;
S^{bg \rightarrow t H^-} 
\left(\alpha_s (\sqrt{s}/{\tilde N})\right) \;
\exp \left[\int_{\sqrt{s}}^{{\sqrt{s}}/{\tilde N}}
\frac{d\mu}{\mu}\; 2 {\rm Re}\Gamma_S^{bg \rightarrow t H^-}
\left(\alpha_s(\mu)\right)\right]
\label{resHS}
\eeqa
where the factorization scale is denoted by $\mu_F$ 
and the renormalization scale by $\mu_R$, and $N$ is the moment 
variable. In the numerical results later we 
will set these two scales equal to each other and denote them 
by $\mu$.
The first exponent in Eq. (\ref{resHS}) is 
\beq
\sum_i E_i(N_i)=
-\sum_i C_i \int^1_0 dz \frac{z^{N_i-1}-1}{1-z}\;
\left \{\int^1_{(1-z)^2} \frac{d\lambda}{\lambda}
\frac{\alpha_s(\lambda s)}{\pi}
+\frac{\alpha_s((1-z)^2 s)}{\pi}\right\}+{\cal O}(\alpha_s^2)
\label{Eexp}
\eeq
with $C_i=C_F=(N_c^2-1)/(2N_c)$ for quarks and 
$C_i=C_A=N_c$ for gluons. 

The second exponent in Eq. (\ref{resHS}) involves the moment-space 
anomalous dimension $\gamma_{i/i}$ of the $\overline{\rm MS}$ parton 
density, 
and the third exponent involves 
the QCD $\beta$ function. $H^{bg \rightarrow tH^-}$ and 
$S^{bg \rightarrow tH^-}$ stand respectively for the 
hard-scattering function and the soft-gluon function.  
$\Gamma_S^{bg \rightarrow tH^-}$ is the soft anomalous dimension, 
and its explicit form at one loop for this process is  
\beq
\Gamma_S^{bg \rightarrow tH^-}=\frac{\alpha_s}{\pi} 
\left[C_F \ln\left(\frac{-t+m_t^2}{m_t\sqrt{s}}
\right)+\frac{C_A}{2} \ln\left(\frac{-u+m_t^2}{-t+m_t^2}\right)
+\frac{C_A}{2} (1-i \pi)\right]+{\cal O} \left(\alpha_s^2\right) \, .
\eeq

We then expand the moment-space expression of Eq. (\ref{resHS}) 
for the resummed cross section through NNNLO and invert  
back to momentum space.
 
The NLO soft gluon corrections can be written as 
\beqa
\frac{d{\hat\sigma}^{(1)}(s_4)}{dt \, du}
&=&F^B
\frac{\alpha_s(\mu_R^2)}{\pi} \left\{
c_{3} \left[\frac{\ln(s_4/m_H^2)}{s_4}\right]_+
+c_{2} \left[\frac{1}{s_4}\right]_+
+c_{1}^{\mu} \, \delta(s_4) \right\}
\label{NLO}
\eeqa
with $F^B$ the Born term, $c_3=2(C_F+C_A)$, 
and expressions for the other coefficients 
as given in \cite{NKcH,NKNNNLO}.

The NNLO soft gluon corrections are \cite{NKcH}
\beq
\frac{d{\hat\sigma}^{(2)}(s_4)}{dt \, du}
=F^B \frac{\alpha_s^2(\mu_R^2)}{\pi^2}
\left\{\frac{1}{2} c_3^2
\left[\frac{\ln^3(s_4/m_H^2)}{s_4}\right]_+
+\left[\frac{3}{2}c_3 \, c_2-\frac{\beta_0}{4} c_3\right] 
\left[\frac{\ln^2(s_4/m_H^2)}{s_4}\right]_+
+\cdots \right\} 
\label{NNLO}
\eeq
where explicit expressions for subleading terms can be found in \cite{NKcH}.

The NNNLO soft gluon corrections are \cite{NKNNNLO}
\beq
\frac{d{\hat\sigma}^{(3)}(s_4)}{dt \, du}
=F^B \frac{\alpha_s^3(\mu_R^2)}{\pi^3}
\left\{\frac{1}{8} c_3^3
\left[\frac{\ln^5(s_4/m_H^2)}{s_4}\right]_+
+\left[\frac{5}{8} c_3^2 \, c_2 -\frac{5}{24} \beta_0 \, c_3^2\right]
\left[\frac{\ln^4(s_4/m_H^2)}{s_4}\right]_+
+\cdots \right\} 
\label{NNNLO}
\eeq
and explicit expressions for the subleading terms and further details 
are given in \cite{NKNNNLO}.

\subsection{$H^-$ production via $bg \rightarrow t H^-$ at the LHC}

We now provide some numerical results for charged
Higgs production in association with a top quark at the LHC.

\begin{figure}
\begin{center}
\epsfig{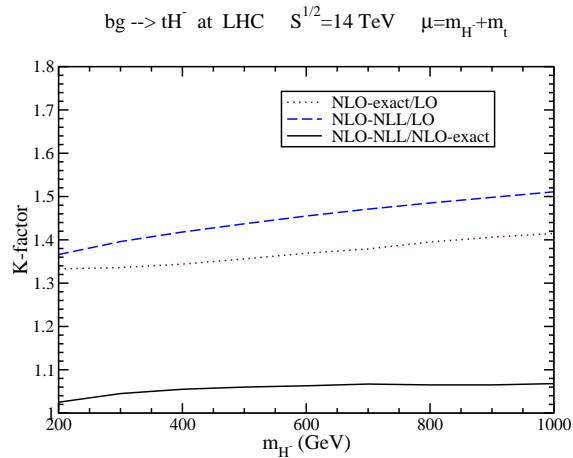}
\caption{NLO exact and approximate $K$ factors for $H^-$ 
production at the LHC.}
\label{higgscompplot}
\end{center}
\end{figure}

We first show that the soft-gluon corrections are dominant by  
comparing the NLO-NLL results with the exact NLO results
that have been derived in Ref. \cite{Zhu01}. We note that
in \cite{Zhu01} the reference scale chosen was 
$\mu=m_H+m_t$.
In most of the results below we will choose $\mu=m_H$, 
which is a natural choice in our calculation.
A cross section known to all orders does not depend on the scale.
However a finite-order cross section does depend on the scale, though the
dependence decreases as we move from LO to NLO, NNLO, NNNLO and so on.
The work in \cite{Zhu01,Plehn02} indeed showed a reduction of scale 
dependence when the NLO corrections are added relative to the LO
cross section.
In fact, as we will see below, the higher-order 
threshold corrections further decrease
the scale dependence, thus resulting in more stable predictions.
But to make the comparison to \cite{Zhu01} we use a scale choice 
$\mu=m_H+m_t$ in Figure \ref{higgscompplot} and 
plot the $K$ factors for $H^-$ production at the LHC. 
The NLO-exact / LO curve shows the enhancement 
from the complete NLO corrections \cite{Zhu01} while the NLO-NLL / LO 
curve shows the contribution of the NLL soft-gluon corrections at NLO.
The two curves are close to each other and this is more easily 
seen from their ratio.
The fact that the NLO-NLL / NLO-exact curve 
is very close to 1 (only a few percent difference) shows 
that the NLO-NLL cross section is a 
remarkably good approximation to the exact NLO result.

\begin{figure}
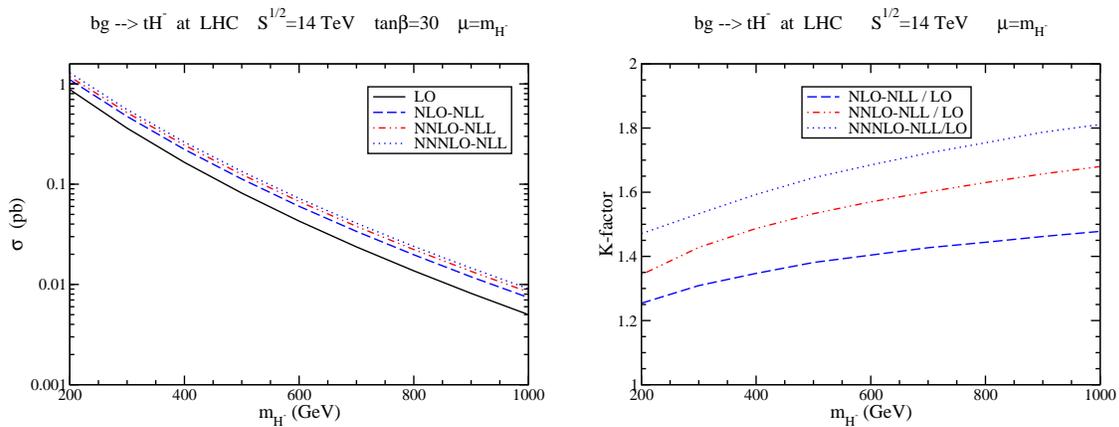

\begin{center}
\epsfig{file=higgs3amhplot.eps,
          height=0.37\textwidth,angle=0}
\hspace{5mm}
\epsfig{file=Khiggs3amhplot.eps,
          height=0.37\textwidth,angle=0}
\caption{The total cross section (left) and $K$ factors (right) 
for $H^-$ production at the LHC.}
\label{higgsmhplot}
\end{center}
\end{figure}

In Figure \ref{higgsmhplot}, on the left, we plot the cross section versus
charged Higgs mass for $pp$ collisions at the LHC with $\sqrt{S}=14$ TeV 
using the MRST2002 approximate 
NNLO parton distributions functions \cite{mrst2002}
with a three-loop evaluation of $\alpha_s$. 
We show results for the LO, NLO-NLL, and NNLO-NLL, and NNNLO-NLL
cross sections, all with a choice of scale $\mu=m_H$  
and with $\tan \beta=30$. 
The cross section spans three orders of magnitude in the mass
range 200 GeV $\le m_H \le$ 1000 GeV.
The higher-order threshold corrections are positive and provide a significant
enhancement to the lowest-order result.
The cross sections for the related process
${\bar b} g \rightarrow {\bar t} H^+$ are exactly the same.  

The right plot of Figure \ref{higgsmhplot} shows the relative size of the 
corrections as $K$ factors at $\mu=m_H$. 
The NLO-NLL / LO curve shows that the
NLO-NLL soft corrections enhance the LO result by 
25\% to 48\% depending on the charged Higgs mass. 
The $K$ factors increase with  
higher masses, as expected, since then we get closer to threshold.
With the NNLO-NLL corrections added we get an enhancement over the LO result 
of 34\% to 68\%.
Adding further the NNNLO-NLL corrections provides an enhancement 
ranging from 47\% to 81\% over the LO result. 

\begin{table}
\begin{center}
\begin{tabular}{|c|c|c|} \hline
\multicolumn{3}{|c|}{$K$ factors}\\ \hline 
$m_H$ (GeV) & {NNLO-NLL} & {NNNLO-NLL} \\ \hline
200  & 1.34 & 1.47  \\ \hline
300  & 1.43 & 1.53  \\ \hline
400  & 1.49 & 1.59  \\ \hline
500  & 1.53 & 1.65  \\ \hline
600  & 1.57 & 1.69  \\ \hline
700  & 1.60 & 1.72  \\ \hline
800  & 1.63 & 1.75  \\ \hline
900  & 1.66 & 1.79  \\ \hline
1000 & 1.68 & 1.81  \\ \hline
\end{tabular}
\caption{The $K$ factors for $H^-$ production at the LHC.}
\end{center}
\end{table}

In Table 1 we show the explicit numbers for the NNLO-NLL and NNNLO-NLL 
$K$ factors for specific values of the charged Higgs mass with $\mu=m_H$.

\begin{figure}
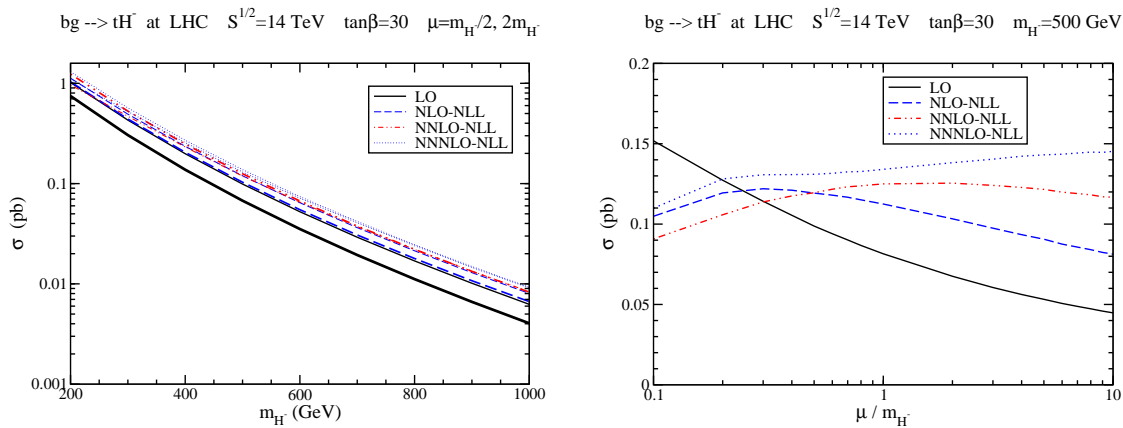

\begin{center}
\epsfig{file=higgs3amhmuplot.eps,
          height=0.37\textwidth,angle=0}
\hspace{5mm}
\epsfig{file=higgs3amucH08plot.eps,
         height=0.37\textwidth,angle=0}
\end{center}
\caption{The scale dependence of the $H^-$ cross section.}
\label{higgsmuplot}
\end{figure}

We next study the scale dependence of the cross section. 
On the left plot of Figure \ref{higgsmuplot} we show the cross 
section at the LHC with $\tan\beta=30$ 
as a function of charged Higgs mass with two different
choices of scale, $\mu=m_H/2$ and $2m_H$.
The scale variation of the LO cross section is quite large. 
The variation at
NLO-NLL is smaller, and at NNLO-NLL and NNNLO-NLL it is very small.
In fact the two NNLO-NLL curves are on top of each other for much
of the range in $m_H$, as are the two NNNLO-NLL curves.

On the right plot of Figure \ref{higgsmuplot}, we show the scale dependence 
of the cross section for $m_H=500$ GeV
and $\tan \beta=30$ over 
a large range in scale, $0.1 \le \mu/m_H \le 10$.
The higher-order threshold corrections greatly decrease the scale dependence 
of the cross section. The NNNLO-NLL curve is relatively flat.
This can also be demonstrated by calculating at each order the ratio of the 
maximum value to the minimal value of the cross section over the $\mu$ range. 
We find 
\begin{tabbing}
$\sigma_{\rm max}/\sigma_{\rm min}=$ \=$ \hspace{5mm} 3.39 \hspace{18mm} 1.50 
\hspace{18mm} 1.38 \hspace{18mm} 1.32$\\
\> \hspace{6mm} $\uparrow \hspace{23mm} \uparrow \hspace{23mm} \uparrow 
\hspace{23mm} \uparrow$
\\
\> \hspace{5mm} LO \hspace{13mm} NLO-NLL \hspace{5mm} NNLO-NLL 
\hspace{5mm} NNNLO-NLL 
\end{tabbing}
We see that with progressing order the ratio decreases and gets closer to one.
 
\begin{figure}
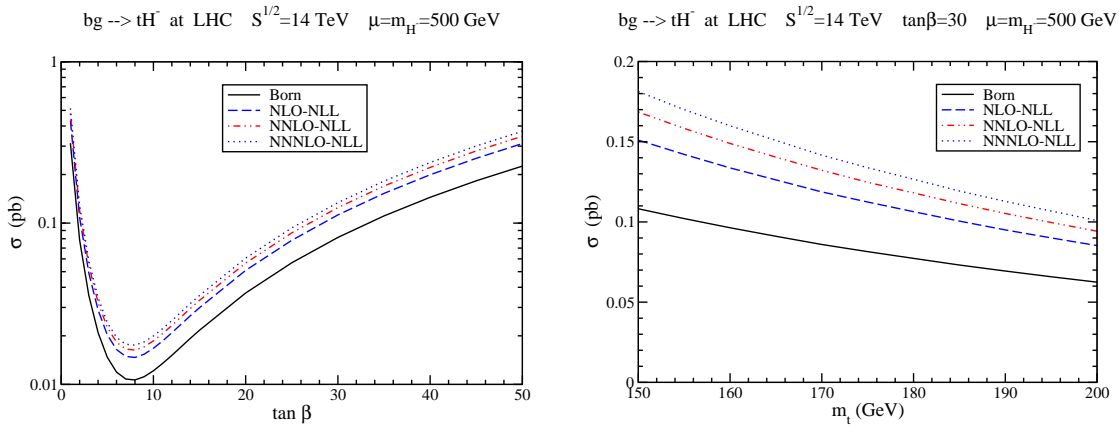

\begin{center}
\epsfig{file=higgs3atanbplot.eps,
          height=0.37\textwidth,angle=0}
\hspace{5mm}
\epsfig{file=higgs3atopplot.eps,
         height=0.37\textwidth,angle=0}
\caption{The $\tan\beta$ (left) and top mass (right) 
dependence of the $H^-$ cross section.}
\label{higgstanbmtplot}
\end{center}
\end{figure}
In Figure \ref{higgstanbmtplot} we plot the dependence of the $H^-$ 
cross section on $\tan \beta$ (left) and the top quark mass (right)
with $\mu=m_H=500$ GeV. 
The $\tan \beta$ variation in the left plot 
is over the range $1 \le \tan \beta \le 50$,
and the cross section is at a minimum near
$\tan\beta=8$. We note that the $\tan \beta$ dependence arises in the
factor $m_b^2 \tan^2 \beta + m_t^2 \cot^2 \beta$,   
and the $\tan\beta$ shape is the same for all curves. 
The dependence on $\tan \beta$ is large, spanning two
orders of magnitude in the range shown.
The dependence of the cross section
on the top quark mass is shown in the right plot for $\tan \beta=30$. 
For heavier top quark masses the cross section decreases.
We see that the dependence is mild
so that the current small experimental uncertainties 
on the top quark mass 
do not play a major role in the total uncertainty of the charged 
Higgs production cross section. 

\section{Other charged Higgs production channels}

We briefly discuss some other production channels for charged Higgs production,
including associated production with a $W$ boson, and pair production.

\subsection{Associated $H^+$ and $W^-$ production}

\begin{figure}
\begin{center}
\begin{picture}(80,120)(0,0)
\ArrowLine(20,50)(0,75)
\ArrowLine(0,25)(20,50)
\ArrowLine(20,50)(60,50)
\DashLine(60,50)(80,25){3}
\DashLine(60,50)(80,75){3}
\Text(0,15)[c]{$b$}\Text(0,85)[c]{${\bar b}$}
\Text(40,65)[c]{$h^0,H^0,A^0$}
\Text(80,15)[c]{$W^-$}\Text(80,85)[c]{$H^+$}
\end{picture}
\hspace{15mm}
\begin{picture}(80,120)(0,0)
\ArrowLine(40,75)(0,75)
\ArrowLine(0,25)(40,25)
\ArrowLine(40,25)(40,75)
\DashLine(40,75)(80,75){3}
\DashLine(40,25)(80,25){3}
\Text(0,15)[c]{$b$}\Text(0,85)[c]{${\bar b}$}
\Text(33,50)[c]{$t$}
\Text(80,15)[c]{$W^-$}\Text(80,85)[c]{$H^+$}
\end{picture}
\caption{LO diagrams for $b {\bar b} \rightarrow H^+ W^-$.}
\label{bbHW}
\end{center}
\end{figure}
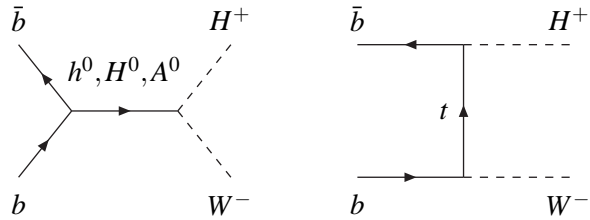

Charged Higgs bosons can be produced in association with $W$ bosons
(see, e.g. \cite{HolZhu01,GLL07,BHK00,EHR06}). 
The LO processes are $gg \rightarrow H^+ W^-$ 
and $b{\bar b} \rightarrow H^+ W^-$.
LO diagrams for  $b{\bar b} \rightarrow H^+ W^-$ are shown 
in Fig. \ref{bbHW}.

Complete NLO calculations for $b {\bar b} \rightarrow H^+ W^-$ were 
presented in \cite{HolZhu01,GLL07}.
In Ref. \cite{EHR06} the above production process followed 
by a leptonic decay of 
$H^+$ and a hadronic decay of $W^-$ was studied as a possibility for 
observing charged Higgs bosons at the LHC.

\subsection{Charged Higgs pair production} 

\begin{figure}
\begin{center}
   \begin{picture}(150,120)(0,0)
        \Gluon(0,60)(40,60){4}{4} \Text(-5,60)[r]{$g$}
        \Vertex(40,60){1}
        \Gluon(0,120)(40,120){4}{4} \Text(-5,120)[r]{$g$}
        \Vertex(40,120){1}
        \ArrowLine(40,120)(40,60) \Text(35,90)[r]{t,b}
        \ArrowLine(70,90)(40,120) \Text(57,107)[lb]{t,b}
        \ArrowLine(40,60)(70,90) \Text(57,73)[lt]{t,b}
        \Vertex(70,90){1}
        \DashLine(70,90)(110,90){5} \Text(90,95)[b]{$h^0,H^0$}
        \Vertex(110,90){1}
        \DashLine(110,90)(140,120){5} \Text(145,120)[l]{$H^+$}
        \DashLine(140,60)(110,90){5} \Text(145,60)[l]{$H^-$}
    \end{picture}
    \hspace{15mm}
    \begin{picture}(150,120)(0,0)
        \Gluon(0,60)(40,60){4}{4}
                      \Text(-5,60)[r]{$g$}
        \Vertex(40,60){1}
        \Gluon(0,120)(40,120){4}{4} \Text(-5,120)[r]{$g$}
        \Vertex(40,120){1}
        \ArrowLine(40,120)(67,93) \Text(49,107)[rt]{t}
        \ArrowLine(73,87)(100,60)
        \ArrowLine(100,60)(40,60) \Text(70,55)[t]{b}
        \ArrowLine(40,60)(100,120) \Text(92,107)[lt]{b}
        \ArrowLine(100,120)(40,120) \Text(70,125)[b]{t}
        \Vertex(100,60){1}
        \Vertex(100,120){1}
        \DashLine(100,120)(140,120){5} \Text(145,120)[l]{$H^-$}
        \DashLine(100,60)(140,60){5} \Text(145,60)[l]{$H^+$}
    \end{picture}

\vspace{10mm}
 
   \begin{picture}(150,60)(0,0)
        \Gluon(5,110)(50,110){4}{4} \Text(0,110)[r]{$g$}
        \Vertex(50,110){1}
        \Gluon(5,60)(50,60){4}{4} \Text(0,60)[r]{$g$}
        \Vertex(50,60){1}
        \ArrowLine(50,110)(50,60) \Text(45,85)[r]{t}
        \ArrowLine(50,60)(100,60) \Text(75,55)[t]{t}
        \ArrowLine(100,60)(100,110) \Text(105,85)[l]{b}
        \ArrowLine(100,110)(50,110) \Text(75,115)[b]{t}
        \Vertex(100,110){1}
        \Vertex(100,60){1}
        \DashLine(100,110)(145,110){5} \Text(150,110)[l]{$H^-$}
        \DashLine(100,60)(145,60){5} \Text(150,60)[l]{$H^+$}
    \end{picture}
    \hspace{15mm}
    \begin{picture}(150,60)(0,0)
        \Gluon(5,110)(50,110){4}{4} \Text(0,110)[r]{$g$}
        \Vertex(50,110){1}
        \Gluon(5,60)(50,60){4}{4} \Text(0,60)[r]{$g$}
        \Vertex(50,60){1}
        \ArrowLine(50,110)(50,60) \Text(45,85)[r]{b}
        \ArrowLine(50,60)(100,60) \Text(75,55)[t]{b}
        \ArrowLine(100,60)(100,110) \Text(105,85)[l]{t}
        \ArrowLine(100,110)(50,110) \Text(75,115)[b]{b}
        \Vertex(100,110){1}
        \Vertex(100,60){1}
        \DashLine(100,110)(145,110){5} \Text(150,110)[l]{$H^+$}
        \DashLine(100,60)(145,60){5} \Text(150,60)[l]{$H^-$}
 \end{picture} 
\vspace{-15mm}
\caption{LO diagrams for $gg \rightarrow H^+ H^-$.}
\label{ggHH}
\end{center}
\end{figure}
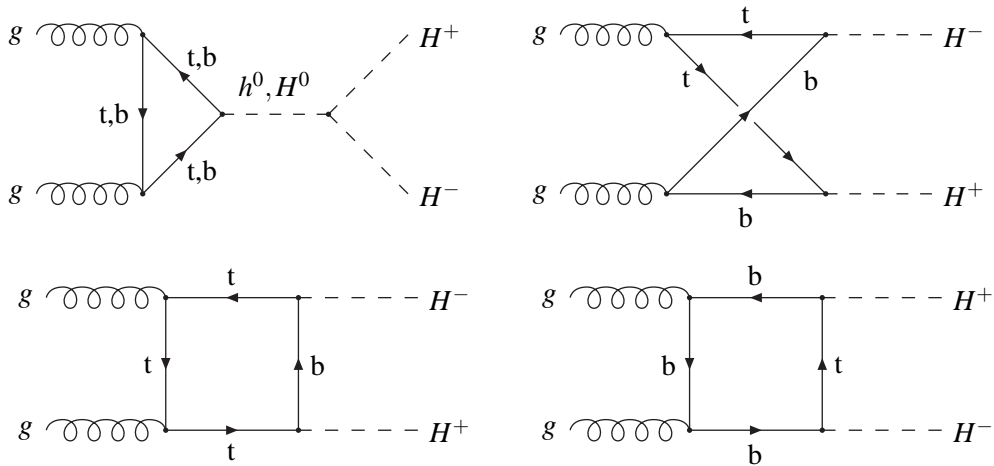

The LO processes for the production of a $H^+$, $H^-$ pair are
$gg \rightarrow H^+ H^-$, $b{\bar b} \rightarrow H^+ H^-$, and 
$q {\bar q} \rightarrow H^+ H^-$ with $q$ a light quark.

LO diagrams for $gg \rightarrow H^+ H^-$, involving loops with top 
and bottom quarks, are shown in Figure \ref{ggHH}. 
Studies have been made for charged Higgs pair production 
via this process at the LHC (see e.g. 
\cite{KPSZ97,Jiang97,BrHol99}). 

\begin{figure}
\begin{center}
\begin{picture}(80,120)(0,0)
\ArrowLine(20,50)(0,75)
\ArrowLine(0,25)(20,50)
\ArrowLine(20,50)(60,50)
\DashLine(60,50)(80,25){3}
\DashLine(60,50)(80,75){3}
\Text(0,15)[c]{$b$}\Text(0,85)[c]{${\bar b}$}
\Text(40,65)[c]{$\gamma,Z,h^0,H^0$}
\Text(80,15)[c]{$H^-$}\Text(80,85)[c]{$H^+$}
\end{picture}
\hspace{30mm}
\begin{picture}(80,120)(0,0)
\ArrowLine(40,75)(0,75)
\ArrowLine(0,25)(40,25)
\ArrowLine(40,25)(40,75)
\DashLine(40,75)(80,75){3}
\DashLine(40,25)(80,25){3}
\Text(0,15)[c]{$b$}\Text(0,85)[c]{${\bar b}$}
\Text(33,50)[c]{$t$}
\Text(80,15)[c]{$H^-$}\Text(80,85)[c]{$H^+$}
\end{picture}
\caption{LO diagrams for $b{\bar b} \rightarrow H^+ H^-$.}
\label{bbHH}
\end{center}
\end{figure}
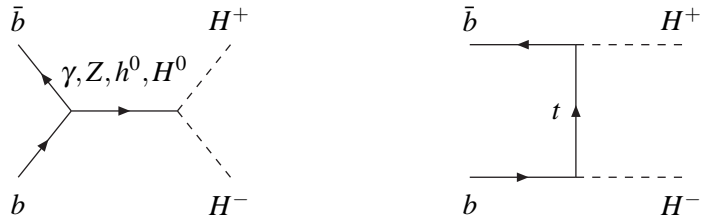

LO diagrams for $b{\bar b} \rightarrow H^+ H^-$ are shown in Figure 
\ref{bbHH}. The NLO corrections for this process have  
been presented in \cite{Hou05,AlPl05}. 
The NLO corrections for the Drell-Yan type process involving light 
quarks, $q{\bar q} \rightarrow H^+ H^-$, 
were also presented in \cite{AlPl05}.
At LO the Drell-Yan type process proceeds via an $s$-channel 
$\gamma$ or $Z$ boson, similar to the left diagram of 
Figure \ref{bbHH}.

The relative contribution of the $gg$, light $q{\bar q}$,  
and $b {\bar b}$ channels to 
charged Higgs pair production at the LHC depends on the values of 
$\tan\beta$ and Higgs mass. At $\tan\beta=10$ the light  $q{\bar q}$ 
contribution is by far dominant but at $\tan\beta=50$ the $gg$ contribution 
dominates \cite{AlPl05}.

Ref. \cite{MorRa03} provides a study of the associated production of 
a charged Higgs pair with a $b{\bar b}$ pair,   
$gg \rightarrow b {\bar b} H^+ H^-$, which is 
the dominant pair production mode at large $\tan\beta$ and 
relevant for the determination of triple-Higgs couplings.

Further calculations of NLO and higher-order corrections will be crucial 
in reducing uncertainties in the theoretical predictions for
cross sections and differential distributions for charged Higgs production.

\end{document}